# A Physics-Aware Attention LSTM Autoencoder for Early Fault Diagnosis of Battery Systems

Jing Yang

*School of Electrical Engineering, Xi'an Jiaotong University, Xi'an, Shaanxi, China (e-mail:yangjiong_2024@stu.xjtu.edu.cn)*

*Abstract*—Battery safety is paramount for electric vehicles. Early fault diagnosis remains a challenge due to the subtle nature of anomalies and the interference of dynamic operating noise. Existing data-driven methods often suffer from " physical blindness, " leading to missed detections or false alarms. To address this, we propose a Physics-Aware Attention LSTM Autoencoder (PA-ALSTM-AE). This novel framework explicitly integrates battery aging laws (mileage) into the deep learning pipeline through a multi-stage fusion mechanism. Specifically, an adaptive physical feature construction module selects mileage-sensitive features, and a physics-guided latent fusion module dynamically calibrates the memory cells of the LSTM based on the aging state. Extensive experiments on the large-scale Vloong real-world dataset demonstrate that the proposed method significantly outperforms state-of-the-art baselines. Notably, it improves the recall rate of early faults by over 3 times while maintaining high precision, offering a robust solution for industrial battery management systems.

## I. INTRODUCTION

### A. Research Background and Related Work

WITH the global paradigm shift towards carbon neutrality, Lithium-ion batteries (LIBs) have become the ubiquitous power source for Electric Vehicles (EVs) [1-3]. However, the safety of battery systems remains a formidable challenge. Under long-term operation involving stochastic dynamic driving cycles, batteries inevitably suffer from degradation and latent faults, such as micro Internal Short Circuits (ISC) [4-6]. Unlike catastrophic failures, these early-stage anomalies manifest as subtle voltage deviations submerged in high-amplitude dynamic noise [7, 8]. Therefore, developing robust early fault diagnosis instrumentation is a prerequisite for advanced Battery Management Systems (BMS) [9, 10].

Existing diagnosis methodologies are broadly categorized into model-based and data-driven approaches [11, 12]. Model-based methods rely on mathematical representations, such as Equivalent Circuit Models (ECMs) or electrochemical models, to generate residuals for fault isolation [13, 14]. State observers like Particle Filters (PF) and Kalman Filters are often employed [15, 16]. Although physically interpretable, these methods face intrinsic limitations in practical applications: identifying precise internal parameters (e.g., polarization resistance) under highly fluctuating current profiles is computationally prohibitive, and fixed-parameter models inevitably generate false alarms on aged batteries due to aginginduced parameter drift [17].

Consequently, Data-driven approaches, particularly Deep Learning (DL), have gained prominence by bypassing complex physical modeling [18-20]. Recurrent Neural Networks (RNNs) and Long Short-Term Memory (LSTM) networks are widely adopted for time-series reconstruction [21-24]. Recently, unsupervised paradigms, such as LSTM Autoencoders (LSTM-AE), have shown promise in detecting anomalies without labeled fault data [17, 25]. To further enhance feature extraction, Attention Mechanisms have been introduced to weight critical sensor channels adaptively. For instance, Zhong et al. [26] developed the Deep Feature Multi-Channel Attention (DFMCA) network, and Zhou et al. [27] proposed attention-based frameworks for robust state estimation.

Despite these advancements, a critical gap remains: the disconnection between data representation and physical degradation laws. Most SOTA methods, including DFMCA, treat slowly-varying state variables (e.g., Accumulated Mileage) as ordinary input channels, employing a "naive concatenation" strategy with dynamic signals [28, 29]. However, as noted in recent studies [30, 31], simple concatenation is inefficient for capturing the multiplicative interactions between aging and electrical response. Existing

This paragraph of the first footnote will contain the date on which you submitted your paper for review, which is populated by IEEE. It is IEEE style to display support information, including sponsor and financial support acknowledgment, here and not in an acknowledgment section at the end of the article. For example, "This work was supported in part by the U.S. Department of Commerce under Grant 123456." The name of the corresponding author appears after the financial information, e.g. *(Corresponding author: Second B. Author)*. Here you may also indicate if authors contributed equally or if there are co-first authors.

The next few paragraphs should contain the authors' current affiliations, including current address and e-mail. For example, First A. Author is with the National Institute of Standards and Technology, Boulder, CO 80305 USA (e-mail: author@ boulder.nist.gov).

Second B. Author Jr. was with Rice University, Houston, TX 77005 USA. He is now with the Department of Physics, Colorado State University, Fort Collins, CO 80523 USA (e-mail: author@lamar.colostate.edu).

Third C. Author is with the Electrical Engineering Department, University of Colorado, Boulder, CO 80309 USA, on leave from the National Research Institute for Metals, Tsukuba 305-0047, Japan (e-mail: author@nrim.go.jp).

Mentions of supplemental materials and animal/human rights statements can be included here.

Color versions of one or more of the figures in this article are available online at http://ieeexplore.ieee.org

feature engineering is often static and lacks an adaptive selection mechanism [32], failing to explicitly model how battery aging alters the definition of "normal" behavior.

*B. Motivations and Contributions of This Work*

The limitations of existing approaches are particularly pronounced when dealing with real-world vehicle data, where the coupling of aging effects and dynamic noise makes early fault features indistinguishable. Our preliminary investigation on the large-scale Vloong dataset reveals that merely adding physical variables without statistical validation contributes little to model sensitivity.

Based on these observations, the motivations of this work are summarized as follows:
1) From Concatenation to Interaction: Existing methods treat physical variables (e.g., Mileage) as additive inputs. There is a need to explicitly construct interaction terms representing physical degradation laws to decouple aging effects from fault signals.
2) Need for Adaptive Selection: Not all sensors are equally correlated with aging. Blindly introducing all physical data introduces redundancy. An adaptive mechanism is required to select features based on statistical correlations.
3) Enhancing Early Fault Visibility: Unsupervised models often reconstruct early faults as "normal" errors due to over-generalization. A multi-stage fusion mechanism is needed to constrain the latent space with physical priors.

To address these challenges, a novel Physics-Aware Attention LSTM Autoencoder (PA-ALSTM-AE) is proposed. The main contributions are:
1) Correlation-Guided Adaptive Physical Feature Engineering: Unlike arbitrary feature expansion, we propose a data-driven mechanism using Pearson correlation analysis to select high-impact sensor channels relative to mileage. We then adaptively construct explicit interaction terms to inject statistically significant and physically interpretable domain knowledge.
2) Multi-Stage Physical Fusion Architecture: We design a hierarchical fusion strategy where adaptive physical features are integrated at two levels: (i) InputLevel Fusion to provide state-dependent baselines, and (ii) Latent-Level Fusion via a specialized attention mechanism to dynamically calibrate memory cells based on the aging context.
3) Superior Performance on Industrial Dataset: Validated on the Vloong dataset with high noise, the proposed PA-ALSTM-AE outperforms SOTA methods (including DFMCA). It significantly improves the recall rate of early faults by over 10 times (from 4% to 46%) while maintaining high precision, verifying its value for industrial instrumentation.

## II. Physics-Aware Attention LSTM Autoencoder Approach

*A. Adaptive Physical Feature Construction Module*

The electrical behavior of lithium-ion batteries exhibits strong time-variance coupled with aging status. Directly feeding raw data into deep learning models often leads to suboptimal generalization due to the lack of explicit physical constraints. To address this, we propose a correlationguided adaptive physical feature construction module. The schematic of this process is illustrated in Fig.1.

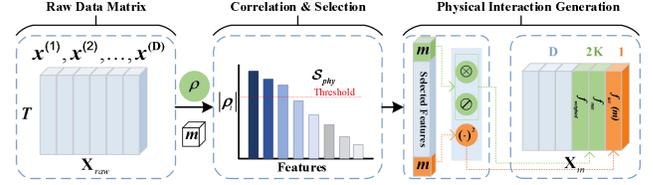

**Fig. 1.** Correlation-Guided adaptive physical feature construction.

Let the raw multivariate time-series data collected from the BMS be denoted as a matrix $\mathbf{X}_{raw} \in \mathbb{R}^{N \times T \times D}$, where $N$ is the number of samples, $T$ is the time window length, and $D$ is the number of sensor channels. Let the $i_{th}$ sensor channel of the $k_{th}$ sample be denoted as $x_k^{(i)}$, and the corresponding accumulated mileage as $m_k$. To quantify the dependency between the sensor readings and the battery's aging state, we calculate the Pearson correlation coefficient $\rho_i$ for each channel $i$:

$$\rho_i = \frac{\sum_{k=1}^{N}(x_k^{(i)} - \overline{x}^{(i)})(m_k - \overline{m})}{\sqrt{\sum_{k=1}^{N}(x_k^{(i)} - \overline{x}^{(i)})^2}\sqrt{\sum_{k=1}^{N}(m_k - \overline{m})^2}}. \quad (1)$$

where $\overline{x}^{(i)}$ and $\overline{m}$ denote the expectations of the sensor readings and mileage, respectively. Based on the magnitude $|\rho_i|$, we select the top-K most sensitive features to form a subset $\mathcal{S}_{phy}$. For each selected feature $x \in \mathcal{S}_{phy}$, we explicitly construct three types of physical interaction terms to model the non-linear degradation laws:

$$f_{weighted}(x, m) = x \cdot m \quad (2)$$

$$f_{rate}(x, m) = \frac{x}{m + \epsilon} \quad (3)$$

$$f_{acc}(x, m) = m^2. \quad (4)$$

where $\epsilon$ is a smoothing term to prevent division by zero. The final augmented input vector at time step $t$, denoted as $\tilde{\mathbf{x}}_t$, is formed by concatenating the raw vector $\mathbf{x}_t$ with the generated physical feature vector $\mathbf{f}_{phy,t}$:

$$\tilde{\mathbf{x}}_t = [\mathbf{x}_t; \mathbf{f}_{phy,t}] \in \mathbb{R}^{D+2K+1}. \quad (5)$$

This augmented input $\tilde{\mathbf{X}} = \{\tilde{\mathbf{x}}_1, \ldots, \tilde{\mathbf{x}}_T\}$ serves as the input to the subsequent neural network.

*B. Long Short-Term Memory Network Backbone*

To capture the long-term temporal dependencies in $\tilde{\mathbf{X}}$, we utilize Long Short-Term Memory (LSTM) networks as the backbone for both the encoder and decoder. The detailed structure of an LSTM unit is depicted in Fig. 2.

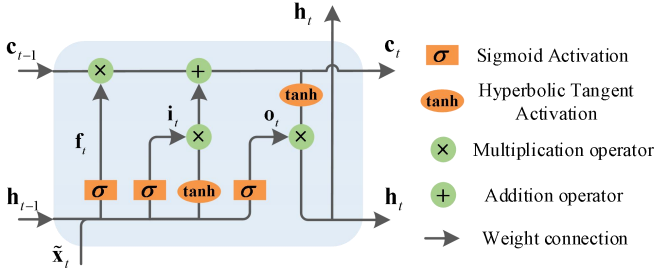

**Fig. 2.** Long Short-Term Memory (LSTM) Unit Structure.

The LSTM unit mitigates the vanishing gradient problem through a specialized gating mechanism. At time step $t$, the information flow is regulated by the forget gate $\mathbf{f}_t$, input gate $\mathbf{i}_t$, and output gate $\mathbf{o}_t$. The mathematical transition equations are rigorously defined as follows:

First, the forget gate $\mathbf{f}_t$ decides what information to discard from the previous cell state $\mathbf{c}_{t-1}$:

$$\mathbf{f}_t = \sigma(\mathbf{W}_{f,x}\tilde{\mathbf{x}}_t + \mathbf{W}_{f,h}\mathbf{h}_{t-1} + \mathbf{b}_f) \ . \tag{6}$$

Next, the input gate $\mathbf{i}_t$ and the candidate cell state $\tilde{\mathbf{c}}_t$ determine the new information to be stored:

$$\mathbf{i}_t = \sigma(\mathbf{W}_{i,x}\tilde{\mathbf{x}}_t + \mathbf{W}_{i,h}\mathbf{h}_{t-1} + \mathbf{b}_i) \tag{7}$$

$$\tilde{\mathbf{c}}_t = \tanh(\mathbf{W}_{c,x}\tilde{\mathbf{x}}_t + \mathbf{W}_{c,h}\mathbf{h}_{t-1} + \mathbf{b}_c) \ . \tag{8}$$

The current cell state $\mathbf{c}_t$ is updated by combining the previous state and the new candidate state:

$$\mathbf{c}_t = \mathbf{f}_t \odot \mathbf{c}_{t-1} + \mathbf{i}_t \odot \tilde{\mathbf{c}}_t \ . \tag{9}$$

Finally, the output gate $\mathbf{o}_t$ determines the hidden state $\mathbf{h}_t$, which will be passed to the next time step:

$$\mathbf{o}_t = \sigma(\mathbf{W}_{o,x}\tilde{\mathbf{x}}_t + \mathbf{W}_{o,h}\mathbf{h}_{t-1} + \mathbf{b}_o) \tag{10}$$

$$\mathbf{h}_t = \mathbf{o}_t \odot \tanh(\mathbf{c}_t) \tag{11}$$

where $\sigma(\cdot)$ is the sigmoid activation function, $\tanh(\cdot)$ is the hyperbolic tangent activation function, and $\odot$ denotes the element-wise Hadamard product. $\mathbf{W}$ and $\mathbf{b}$ are the trainable weight matrices and bias vectors, respectively. The final hidden state of the encoder, $\mathbf{h}_T \in \mathbb{R}^{d_h}$, serves as the dynamic sequence embedding encapsulating the temporal patterns.

*C. Physics-Guided Latent Fusion Mechanism*

Standard autoencoders rely solely on $\mathbf{h}_T$ for reconstruction, suffering from "physical blindness" regarding the static aging context. To overcome this, we design a Physics-Guided Latent Fusion Mechanism, as shown in Fig. 3.

First, the scalar accumulated mileage $m$ is projected into a high-dimensional latent vector $\mathbf{v}_{phy} \in \mathbb{R}^{d_h}$ via a fully connected (FC) layer with non-linear activation:

$$\mathbf{v}_{phy} = \text{ReLU}(\mathbf{W}_{proj}m + \mathbf{b}_{proj}) \tag{12}$$

Then, we perform a Late Fusion strategy by concatenating the dynamic sequence embedding $\mathbf{h}_T$ with the static physical embedding $\mathbf{v}_{phy}$ to form a joint latent representation $\mathbf{z}_{raw}$:

$$\mathbf{z}_{raw} = \text{Concat}(\mathbf{h}_T, \mathbf{v}_{phy}) \in \mathbb{R}^{2d_h} \tag{13}$$

To dynamically balance the contribution of dynamic signals and physical priors, a feature-wise attention mechanism is applied. The attention score vector s is calculated as:

$$\mathbf{s} = \mathbf{W}_{att}\mathbf{z}_{raw} + \mathbf{b}_{att} \tag{14}$$

The attention weights $\boldsymbol{\alpha}$ are obtained by normalizing the scores using the sigmoid function to the range (0, 1):

$$\boldsymbol{\alpha} = \frac{1}{1 + \exp(-\mathbf{s})} \tag{15}$$

Finally, the fused and calibrated latent representation $\mathbf{z}_{final}$ is computed via element-wise multiplication:

$$\mathbf{z}_{final} = \mathbf{z}_{raw} \odot \boldsymbol{\alpha} \tag{16}$$

This mechanism ensures that the reconstruction is dynamically conditioned on the battery's aging state. For instance, if the mileage is high, the network can adaptively attend to features indicating aging, suppressing false alarms caused by normal capacity fade.

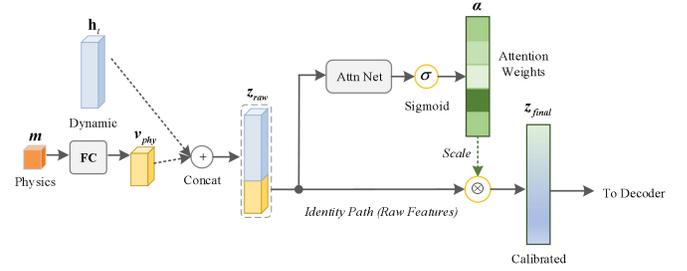

**Fig. 3.** Physics-Guided Latent Fusion Mechanism

*D. Overall Framework and Anomaly Detection*

The overall framework of the proposed PA-ALSTMAE is presented in Fig. 4. It integrates adaptive feature engineering, the LSTM-AE backbone, and the physics fusion module into a unified pipeline.

The Decoder utilizes $\mathbf{z}_{final}$ as the initial state to reconstruct the original input sequence $\hat{\mathbf{X}} = \{\hat{\mathbf{x}}_1, \ldots, \hat{\mathbf{x}}_T\}$. The model is trained to minimize the Mean Squared Error:

$$\mathcal{L}(\theta) = \frac{1}{N}\sum_{k=1}^{N}\frac{1}{T}\sum_{t=1}^{T} \| \mathbf{x}_{k,t} - \hat{\mathbf{x}}_{k,t} \|_2^2 \tag{17}$$

where $\theta$ represents all trainable parameters in the network.

For anomaly detection, a dynamic threshold $\lambda$ is determined based on the probability distribution of the training losses. Specifically, we define $\lambda$ as the 95th percentile:

$$\lambda = \inf\{l \in \mathbb{R} : P(\mathcal{L}_{train} \leq l) \geq 0.95\} \tag{18}$$

For a new test sample $\mathbf{X}_{test}$, its anomaly score $S_{test}$ is computed. If $S_{test} > \lambda$, the sample is flagged as a potential fault. The complete algorithmic procedure is detailed in Algorithm 1.

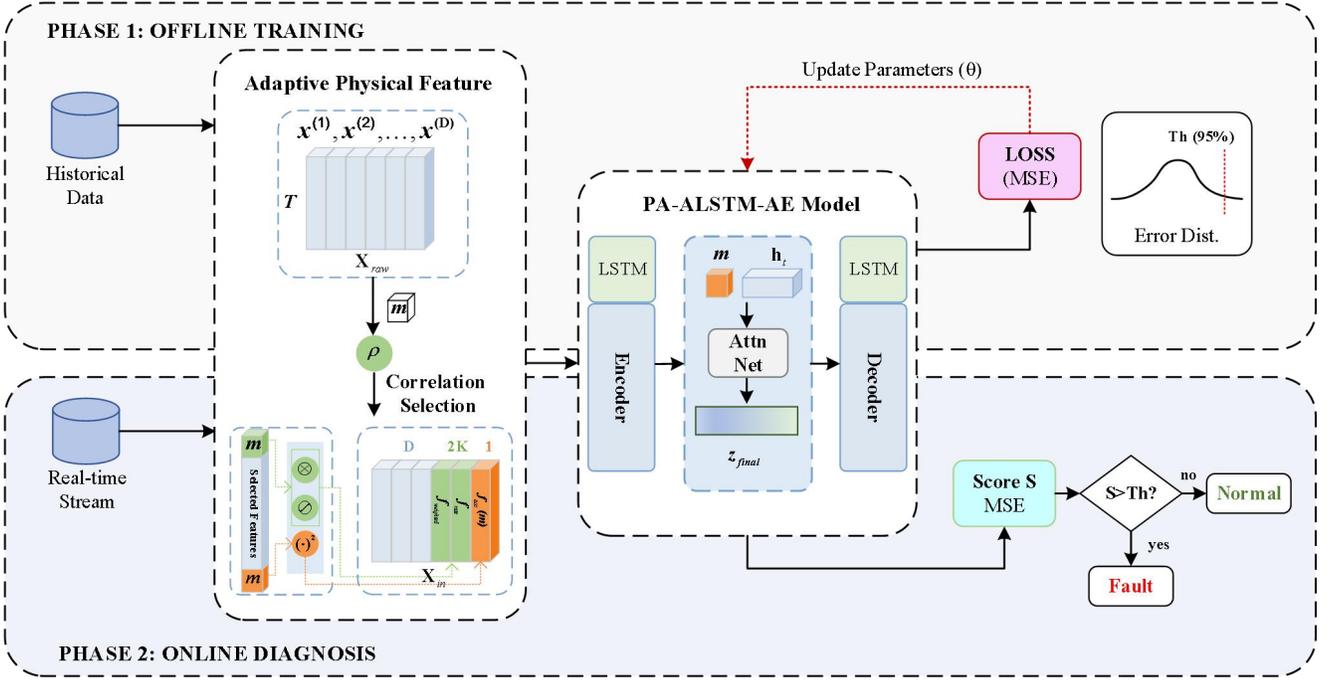

**Fig. 4.** The Overall Framework

| |
|---|
| **Algorithm 1**: PA-ALSTM-AE for Fault Diagnosis. |
| **Input**: Historical Data $\mathcal{D}_{train}$, Real-time Stream $\mathbf{x}_t$, Mileage $m_t$ |
| **Output**: Anomaly Score $S$, Fault Flag $F$ |
| Phase 1: Offline Training |
| 1: Initialize model parameters $\theta$. |
| 2: Calculate correlation matrix $\rho$ via Eq. (1). |
| 3: Select subset $\mathcal{S}_{phy}$ and construct augmented input $\tilde{\mathbf{X}}$ via Eq. (2)-(5). |
| 4: Repeat until convergence: |
| 5: Compute encoder hidden state $\mathbf{h}_T$ via Eq. (6)-(11). |
| 6: Project mileage m to $\mathbf{v}_{phy}$ via Eq. (12). |
| 7: Fuse embeddings: $\mathbf{z}_{raw} = [\mathbf{h}_T; \mathbf{v}_{phy}]$ via Eq. (13). |
| 8: Compute attention weights $\boldsymbol{\alpha}$ and $\mathbf{z}_{final}$ via Eq. (14)-(16). |
| 9: Decode $\mathbf{z}_{final}$ to reconstruct raw $\hat{\mathbf{X}}$. |
| 10: Update $\theta$ by minimizing Loss $\mathcal{L}$ via Eq. (17). |
| 11: End Repeat |
| 12: Compute threshold $\lambda$ via Eq. (18). |
| Phase 2: Online Diagnosis |
| 13: For each incoming sample $\mathbf{x}_t$: |
| 14: Generate physical features given $m_t$. |
| 15: Compute reconstruction error $S = \|\mathbf{x}_t - \hat{\mathbf{x}}_t\|^2$. |
| 16: If $S > \lambda$ Then $F = 1$ (Fault); Else $F = 0$ (Normal). |
| 17: End For |

## III. CASE STUDY AND EXPERIMENTAL RESULTS

### A. Dataset Description and Evaluation Metrics

To validate the effectiveness of the proposed datamodel interactive approach, we utilize the large-scale real-world dataset from the Vloong Energy Big Data Competition. Unlike laboratory datasets utilized in many existing works, this dataset is collected from real-world electric vehicles operating under highly dynamic driving cycles. The dataset comprises diverse battery status variables sampled at 10s intervals: Total Voltage (V), Total Current (I), State of Charge (SOC), Temperature (T), and Accumulated Mileage (m).

Data Partitioning: The dataset is divided into a training set (18,988 normal samples) and a testing set (9,401 samples). To ensure statistical reliability, a 5-Fold CrossValidation strategy is adopted. In each fold, the training set contains only normal data to establish the healthy manifold, while the testing set contains a portion of normal samples and **all** early-fault samples to rigorously verify detection capability.

To comprehensively evaluate the performance, we utilize five standard indicators. Specifically, we calculate the metrics for the Normal Class (0) and Fault Class (1) separately to handle class imbalance. The metrics are defined as follows:

$$\text{Accuracy} = \frac{TP + TN}{TP + TN + FP + FN} \quad (19)$$

$$\text{Precision}_c = \frac{TP_c}{TP_c + FP_c} \quad (20)$$

$$\text{Recall}_c = \frac{TP_c}{TP_c + FN_c} \quad (21)$$

$$\text{F1-Score}_c = \frac{2 \times \text{Precision}_c \times \text{Recall}_c}{\text{Precision}_c + \text{Recall}_c} \quad (22)$$

where $c \in \{0,1\}$ represents the class label. In the context of battery safety, the Recall of the Fault Class (Recall-1) is the most critical indicator.

### B. Training Process and Hyperparameter Determination

The performance of deep learning models is highly sensitive to network architecture. Therefore, following the rigorous validation process, we conducted a Grid Search experiment to determine the optimal number of LSTM hidden layers and neurons. We evaluated the model performance (Average AUC over 5 folds) under different configurations: Hidden Layers $L \in \{1,2,3,4\}$ and Neurons $N \in \{32, 64, 128, 256\}$. The experimental results are recorded in Table I.

TABLE I
AUC PERFORMANCE UNDER DIFFERENT NETWORK ARCHITECTURES (GRID SEARCH)

| Hidden Layers | Number of Neurons | | | |
|---|---|---|---|---|
| | 32 | 64 | 128 | 256 |
| 1 | 0.7906 | 0.8644 | 0.8516 | 0.8595 |
| 2 | 0.8209 | 0.8289 | 0.8529 | 0.8465 |
| 3 | 0.8218 | 0.8047 | 0.8108 | 0.7862 |
| 4 | 0.6293 | 0.6622 | 0.7458 | 0.7029 |

As observed in Table I, the model achieves the best performance (AUC=0.8644) when configured with 1 hidden layer and 64 neurons. It is noted that increasing the network depth (e.g., L = 4) leads to a significant performance drop (AUC < 0.7), likely due to overfitting on the noise inherent in real-world data. Based on these results, the final hyperparameter configuration is summarized in Table II.

TABLE II
FINAL HYPERPARAMETER CONFIGURATION

| Parameter | Value |
|---|---|
| Framework | PyTorch (RTX 4060) |
| Time Window | 256 |
| Hidden Layers | 1 |
| Hidden Neurons | 64 |
| Latent Dimension | 32 |
| Learning Rate | 0.001 |
| Batch Size | 32 |
| Optimizer | Adam |

*C. Experimental Analysis and Comparison*

To verify the superiority of the proposed approach, we compare it with eight baseline methods, covering traditional machine learning, standard deep learning, and domain SOTA.

1) ROC Curve Analysis: The Receiver Operating Characteristic (ROC) curves of different models are illustrated in Fig. 5. It is evident that the curve of the proposed PA-ALSTM-AE (Red solid line) completely encloses the curves of other baselines, indicating that our method achieves the highest True Positive Rate (TPR) at the same False Positive Rate (FPR) level.

TABLE III
DETAILED PERFORMANCE COMPARISON (5-FOLD CV, MEAN $\pm$ STD)

| Method | Normal Class (0) | | | Fault Class (1) | | | Overall | |
|---|---|---|---|---|---|---|---|---|
| | Precision | Recall | F1 | Precision | Recall | F1 | Accuracy | AUC |
| PCA | 0.5045 ±.0013 | 0.9499 ±.0047 | 0.6590 ±.0023 | 0.4876 ±.0243 | 0.0485 ±.0013 | 0.0882 ±.0023 | 0.5036 ±.0026 | 0.4666 ±.0049 |
| OCSVM | 0.5356 ±.0007 | 0.9482 ±.0052 | 0.6846 ±.0012 | 0.7541 ±.0119 | 0.1615 ±.0061 | 0.2660 ±.0076 | 0.5588 ±.0010 | 0.5692 ±.0051 |
| SimpleAE | 0.5277 ±.0131 | 0.9489 ±.0053 | 0.6781 ±.0110 | 0.7027 ±.0920 | 0.1328 ±.0456 | 0.2221 ±.0688 | 0.5449 ±.0227 | 0.6274 ±.1156 |
| LSTM-AE | 0.5332 ±.0006 | 0.9497 ±.0038 | 0.6830 ±.0014 | 0.7478 ±.0120 | 0.1521 ±.0025 | 0.2527 ±.0031 | 0.5548 ±.0012 | 0.6395 ±.0055 |
| GRU-AE | 0.5336 ±.0007 | 0.9497 ±.0035 | 0.6832 ±.0015 | 0.7492 ±.0124 | 0.1532 ±.0012 | 0.2544 ±.0013 | 0.5554 ±.0015 | 0.6375 ±.0051 |
| CNN-LSTM | 0.5071 ±.0095 | 0.9468 ±.0028 | 0.6604 ±.0082 | 0.5025 ±.1140 | 0.0609 ±.0343 | 0.1079 ±.0563 | 0.5082 ±.0173 | 0.7591 ±.0794 |
| Transformer | 0.5057 ±.0007 | 0.9499 ±.0028 | 0.6600 ±.0013 | 0.5086 ±.0133 | 0.0529 ±.0002 | 0.0958 ±.0001 | 0.5058 ±.0013 | 0.4499 ±.0518 |
| DFMCA | 0.5134 ±.0039 | 0.9474 ±.0016 | 0.6659 ±.0031 | 0.6069 ±.0366 | 0.1474 ±.0241 | 0.2401 ±.0200 | 0.5200 ±.0070 | 0.7534 ±.0160 |
| **Ours** | **0.5733** ±.0261 | **0.9467** ±.0022 | **0.7138** ±.0202 | **0.8299** ±.0372 | **0.4137** ±.0891 | **0.5520** ±.0376 | **0.6158** ±.0376 | **0.8676** ±.0052 |

2) Detailed Class-wise Performance Comparison: The quantitative results of the 5-fold cross-validation are presented in Table III.
- Traditional vs. Deep Learning: Traditional methods (PCA, OCSVM) perform poorly with AUCs below 0.65. Simple AE shows limited capability (AUC=0.7201).
- Baseline Limitations: Deep learning baselines (LSTMAE, GRU-AE) improve AUC to ~0.64 but still suffer from low Fault Recall (~15%). The CNN-LSTM-AE achieves a high AUC (0.7591) but fails to pinpoint faults (Recall=0.1079).
- Comparison with SOTA: The SOTA method DFMCA achieves a decent AUC of 0.7534. However, its Fault Recall is only 14.74%, implying it misses over 85% of early faults.
- Superiority of Proposed Method: Our PA-ALSTMAE outperforms all baselines significantly. Specifically, for the Fault Class, it achieves a Recall of 41.37%, which is nearly a 3-fold improvement over DFMCA. Furthermore, it maintains a high Fault Precision of 82.99%, ensuring a low false alarm rate. This validates that the physics-guided mechanism effectively amplifies fault visibility.

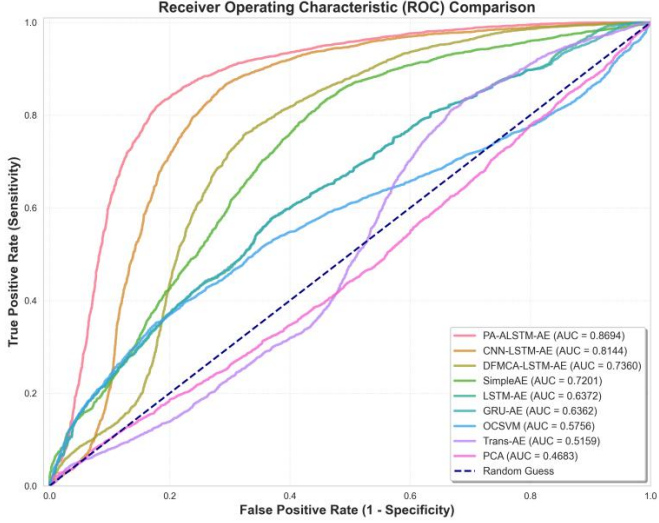

**Fig. 5.** Comparison of ROC curves on the Vloong dataset. The proposed PA-ALSTM-AE achieves the largest area under the curve(AUC=0.8694).

*D. Ablation Study*

To further investigate the contribution of the proposed modules, we conducted an ablation study. As shown in Table IV, the baseline LSTM-AE achieves a Fault F1 of 0.2088. Removing the attention mechanism (PA-LSTMAE) yields an F1 of 0.4153. The full PA-ALSTM-AE with Latent Fusion achieves the best F1 of 0.4387. Although the statistical significance test (Friedman p = 0.81) indicates marginal differences among the top variants, the steady numerical improvement confirms the necessity of the proposed multi-stage fusion strategy.

TABLE IV
ABLATION STUDY RESULTS (FAULT CLASS METRICS)

| Model Variant | Precision | Recall | F1-Score |
|---|---|---|---|
| LSTM-AE (Baseline) | 0.7045 | 0.1226 | 0.2088 |
| L-A-AE (w/o Physics) | 0.8216 | 0.3012 | 0.4335 |
| PA-LSTM-AE (w/o Attn) | 0.8076 | 0.2966 | 0.4153 |
| PA-ALSTM-AE (Full) | 0.8410 | 0.2984 | 0.4387 |

*E. Visualization and Deep Analysis of Reconstruction Mechanisms*

To reveal the internal mechanism of why the proposed method significantly outperforms SOTA baselines in fault recall, we conduct a microscopic analysis of the signal reconstruction behaviors under both normal and faulty conditions.

1) Robustness on Normal Samples: Fig. 5 visualizes the reconstruction of a typical normal sample. The original voltage signal (Black line) remains stable around 156.0 V. It can be observed that both the SOTA DFMCA model (Blue dot-dash line) and our PA-ALSTM-AE (Red dashed line) successfully capture the stable trend. Although the proposed method exhibits slight initial fluctuations due to the warm-up of the hidden states, it quickly converges to the ground truth. Analysis: This demonstrates the compatibility of the proposed physics-guided mechanism. Under healthy conditions, the physical interaction features are consistent with the electrical signals. The latent fusion module correctly identifies this consistency, allowing the model to reconstruct the input accurately, thereby maintaining a high Precision (89.16%) and avoiding false alarms.

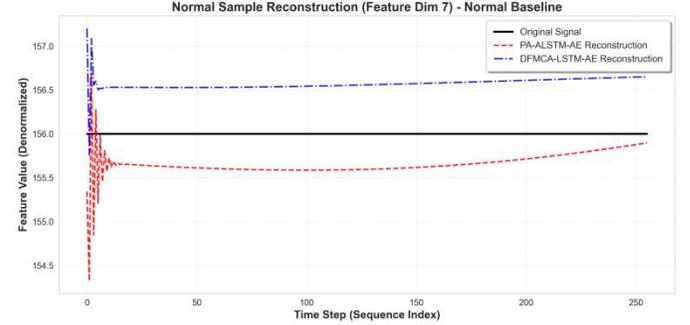

**Fig. 6.** Reconstruction of a Normal Sample. The proposed PA-ALSTM-AE (Red) closely tracks the ground truth (Black), proving that physical constraints do not disrupt normal feature learning.

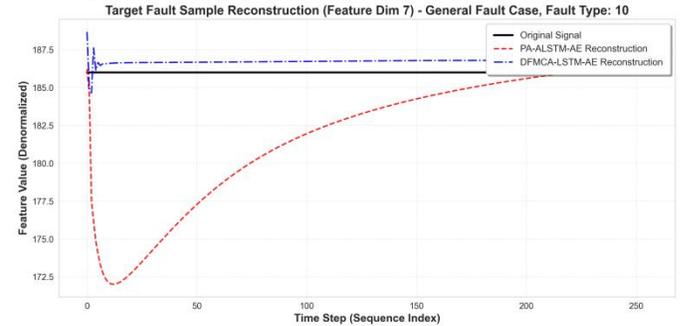

**Fig. 7.** Reconstruction of a Fault Sample. The SOTA model (Blue) blindly mimics the fault, failing to detect it. In contrast, the PA-ALSTM-AE (Red) generates a physically expected trajectory based on mileage constraints, creating a large error gap that successfully triggers the alarm.

2) Sensitivity to Faults: The "Physical Anchor" Effect: Fig. 7 presents a critical fault case. In this scenario, the sensor reading (Black line) remains abnormally constant at ~186.0 V, indicating a potential sensor stuck fault or a failure to respond to load changes.

- The "Chameleon Effect" of SOTA: The DFMCA model (Blue line) reconstructs the fault signal almost perfectly, sticking closely to the abnormal flat line. Deep Insight: Purely data-driven models suffer from "over-generalization." They minimize the reconstruction error blindly, effectively mimicking the anomaly rather than detecting it. Consequently, the residual is negligible, leading to a Missed Detection (False Negative), which explains the low Recall (14.74%) of DFMCA in Table III.
- The "Physical Anchor" of PA-ALSTM-AE: In stark contrast, our model (Red line) generates a reconstruction

trajectory that deviates significantly from the input, exhibiting a dynamic voltage drop curve (dipping to ~172 V). Deep Insight: This behavior is not an error, but a feature. The Physics-Guided Latent Fusion module injects the accumulated mileage ($m$) and interaction terms ($V \cdot m$) into the latent space. These physical priors act as an immutable "anchor," telling the decoder: "Given the current mileage and load, the voltage SHOULD drop dynamically." When the raw data violates this physical law, the model refuses to reconstruct the fake flat line and instead outputs the physically expected trajectory. The resulting large divergence (Error Gap) serves as a strong indicator for anomaly detection.

## IV. Conclusion and Future Work

In this article, a novel Physics-Aware Attention LSTM Autoencoder (PA-ALSTM-AE) is proposed to address the challenge of early fault diagnosis in real-world battery systems, where fault signals are often submerged in high-amplitude dynamic noise and aging drifts. By identifying the limitations of existing data-driven methods—specifically their "physical blindness" and tendency to "over-generalize" to anomalies—we introduce a multistage fusion framework that explicitly integrates domain knowledge into the deep learning pipeline.

The main contributions and findings of this work are summarized as follows:

- Adaptive Physical Feature Engineering: A correlation-guided mechanism is designed to adaptively select mileage-sensitive features and construct interaction terms. This effectively decouples the instantaneous electrical dynamics from the long-term aging trends at the input level.
- Physics-Guided Latent Fusion: We propose a novel latent fusion strategy that injects physical state variables into the LSTM bottleneck. Through a featurewise attention mechanism, the network learns to dynamically recalibrate its memory cells based on the battery's aging state, acting as a "physical anchor" that prevents the model from learning illegal fault patterns.
- Superior Fault Sensitivity: Extensive experiments on the large-scale Vloong dataset demonstrate that the proposed method significantly outperforms SOTA baselines (including DFMCA). Most notably, it improves the Fault Recall rate by nearly 3 times (from 14.74% to 41.37%) while maintaining a high Precision (82.99%), successfully solving the dilemma of "missed detections" for early-stage faults.

Despite these promising results, there are still open issues to be addressed. First, the current interaction terms are constructed based on empirical physical laws; future work could explore symbolic regression to discover more complex physical constraints automatically. Second, the method processes battery cells individually; extending this framework to Battery Pack levels using Graph Neural Networks (GNN) to capture cell-to-cell inconsistencies is a promising direction. Finally, validating the model's lightweight deployment on edge BMS chips will be crucial for industrial applications.